\documentclass[pmlr]{jmlr}% new name PMLR (Proceedings of Machine Learning)

\RequirePackage{graphicx}
 \usepackage{booktabs}
\usepackage{longtable}% for long tables
\makeatletter
\def\set@curr@file#1{\def\@curr@file{#1}} %temp workaround for 2019 latex release
\makeatother
\usepackage[load-configurations=version-1]{siunitx} % newer version

\theorembodyfont{\upshape}
\theoremheaderfont{\scshape}
\theorempostheader{:}
\theoremsep{\newline}

\jmlrvolume{298}
\jmlryear{2026}
\title[Heterogeneity Detection with Likelihood Clustering]{Detecting Batch Heterogeneity via Likelihood Clustering}

\author{\Name{Austin Talbot}
       \Email{talbota@pillarbiosci.com}\\ 
       \addr Pillar Biosciences Inc\\
       Natick, MA, USA 
       \AND
       \Name{Yue Ke}
       \Email{key@pillarbiosci.com}\\ 
       \addr Pillar Biosciences Inc\\
       Natick, MA, USA 
       }

\begin{document}

\maketitle

\begin{abstract}

Batch effects represent a major confounder in genomic diagnostics. In copy number variant (CNV) detection from next-generation sequencing, many algorithms compare read depth between test samples and a reference derived from the processing batch, assuming samples are process-matched. When this assumption is violated, with causes ranging from reagent lot changes and sample quality differences to multi-site processing, the reference becomes inappropriate, introducing false CNV calls or masking true pathogenic variants. Detecting such heterogeneity before downstream analysis is critical for reliable clinical interpretation. Existing batch effect detection methods either cluster samples based on raw features, risking conflation of biological signal with technical variation, or require known batch labels that are frequently unavailable. We introduce a method that addresses both limitations by clustering samples according to their Bayesian model evidence. The central insight is that evidence quantifies compatibility between data and model assumptions, technical artifacts violate assumptions and reduce evidence, whereas biological variation, including CNV status, is anticipated by the model and yields high evidence. This asymmetry provides a discriminative signal that separates batch effects from biology. We formalize heterogeneity detection as a likelihood ratio test for mixture structure in evidence space, using parametric bootstrap calibration to ensure conservative false positive rates. We validate our approach on synthetic data demonstrating proper Type I error control, three clinical targeted sequencing panels (liquid biopsy, BRCA, and thalassemia) exhibiting distinct batch effect mechanisms, and mouse electrophysiology recordings demonstrating cross-modality generalization. Our method achieves superior clustering accuracy compared to standard correlation-based and dimensionality-reduction approaches while maintaining the conservativeness required for clinical usage.

\end{abstract}

\section{Introduction}\label{sec1}
Batch effects are a persistent source of error in genomic
diagnostics~\citep{leek2010tackling,johnson2007adjusting}. In copy number variant (CNV) detection from next-generation sequencing (NGS), many methods compare read depth between a test sample and a reference derived from the processing batch~\citep{Fromer2012XHMM,krumm2012copy,alkan2011genome}. This approach assumes that samples within a batch are 
process-matched, that is they are prepared under identical conditions, sequenced with the same reagents on the same instrument run, and analyzed through the same bioinformatics pipeline, so that the reference captures technical variation unrelated to copy number. When this assumption is violated, the batch reference becomes inappropriate, introducing spurious CNV  calls that may trigger unnecessary invasive procedures or, conversely, mask true pathogenic variants that would otherwise prompt clinical intervention~\citep{risso2014normalization,kusmirek2019comparison}. Detecting such heterogeneity before downstream analysis is performed is  critical for reliable diagnostics.

Existing approaches to batch effect detection fall broadly into two categories. Unsupervised methods cluster samples based on raw features, such as principal components of read counts or coverage profiles, and flag clusters that correlate with technical covariates~\citep{leek2007capturing}. However, these methods are prone to confounding biological signal with technical variation: a cluster of samples carrying true CNVs may be misidentified as a batch artifact, leading to unnecessary re-processing or exclusion of valid data. Supervised methods assess the association between known batch labels and molecular measurements~\citep{johnson2007adjusting}, but in clinical sequencing workflows, batch membership is often unknown or incompletely recorded. 

A practical heterogeneity detection method must satisfy several
competing criteria. First, it must be conservative: in clinical settings, flagging a batch as heterogeneous triggers costly re-sequencing or manual review, so false alarms carry real operational burden. Second, it must be sensitive to diverse technical effects, not just systematic shifts in coverage, but also increased variance, outlier amplicons, or subtle changes in noise structure that manifest differently across samples. Third, and most critically, it must distinguish technical artifacts from true biological signal: a method that flags batches simply because they contain CNV-positive samples would be counterproductive, as CNVs are precisely what we aim to detect. Fourth, it must be robust to individual outliers, a single failed sample should be flagged as such, not interpreted as evidence of global batch structure. Satisfying all four criteria simultaneously is challenging: sensitive methods tend to over-detect structure, conservative methods miss subtle effects, and standard clustering approaches cannot distinguish technical from biological heterogeneity.

We introduce a heterogeneity detection method that addresses these challenges by clustering samples according to their Bayesian model evidence (marginal likelihood), which quantifies how well the assumed statistical model explains each sample's data~\citep{kass1995bayes,mackay2003information}. The key insight is that model evidence measures compatibility between data and model assumptions, not merely goodness of fit to a point estimate. A CNV detection model is designed to explain samples with copy number changes; true CNV-positive samples are well-modeled and yield high evidence. In contrast, technical artifacts, such as reagent degradation, pipetting errors, or reference mismatch, which introduce noise patterns that the model does not anticipate, violate its assumptions and systematically reduce evidence. This asymmetry allows evidence to serve as a discriminative signal as low evidence indicates unexpected behavior (assumption violations), while biological variation produces expected behavior that the model handles well. By examining the distribution of evidence values across a cohort rather than clustering on raw features, we isolate the signal of technical heterogeneity from the signal of biological variation. 

In this work, we demonstrate our method on synthetic data as well as three different assays, each detecting a distinct source of batch effects. Furthermore, we demonstrate that our method is broadly applicable, detecting batch effects in mouse electrophysiology. We show that our method reliably detects a wide range of sources of heterogeneity, while avoiding false positives when such effects are known to not exist. We also show that the model evidence outperforms alternative methods for distinguishing the groups when batch effects are present.

The contents of this work are as follows. In Section~\ref{sec2}, we summarize related work for detecting and characterizing batch effects. In Section~\ref{sec3}, we describe our method, particularly as applied to the task of CNV detection. Section~\ref{sec4} provides an analysis of our method on synthetic data, demonstrating that our test is conservative. Section~\ref{sec3p2} describes the cohorts on which our method is evaluated, along with feature and processing descriptions. We then analyze these datasets in Section~\ref{sec5}. Finally, in Section~\ref{sec6} we provide brief concluding remarks and directions for future research. Python code implementing all models, loading the data used, and scripts required to reproduce all results is available at \url{https://github.com/Pillar-Biosciences-Inc/BatchDetect}.

\subsection*{Generalizable Insights about Machine Learning in the Context of Healthcare}
\begin{itemize}
    \item \textbf{Evidence as a diagnostic signal.} Bayesian model evidence provides a principled, label-agnostic measure of assumption violation. Unlike prediction error or reconstruction loss, evidence integrates over parameter uncertainty and naturally penalizes model misspecification, enabling heterogeneity detection without batch labels and without misclassifying biologically positive samples as artifacts.
    \item \textbf{Distribution shift detection in evidence space.} Framing data quality monitoring as distribution shift detection over model fit as opposed to input features sidesteps the fundamental challenge of disentangling technical from biological variation, a problem that confounds many alternative approaches.
    \item \textbf{Calibrated hypothesis testing for structure detection.} The parametric bootstrap framework for testing mixture structure is broadly applicable: any clinical ML pipeline that computes per-sample likelihoods can incorporate this approach to monitor for batch effects or data drift, with rigorous false discovery control.
\end{itemize}

\section{Related Work}\label{sec2}
Batch effects have long been recognized as a major confounder in genomic analyses~\citep{leek2010tackling}. In CNV detection, several methods mitigate batch-specific noise through dynamic reference selection. ExomeDepth~\citep{plagnol2012robust} constructs reference panels from samples with maximal read-depth correlation; CLAMMS~\citep{packer2016clamms} and CANOES~\citep{backenroth2014canoes} employ similar strategies. These approaches can substantially improve CNV calling accuracy, particularly in targeted panels where sample-to-sample variability is high~\citep{kusmirek2019comparison}. However, reference-selection methods operate reactively—they optimize the reference for each sample but do not explicitly identify whether batch heterogeneity exists. When unrecognized subgroups are present, correlation-based selection may still construct inappropriate references, propagating systematic errors. Grouping samples by known batch labels (e.g., library preparation date) prior to calling can reduce false positives~\citep{krumm2012copy}, but this requires metadata that is not always available or reliable.

In contrast to the mature literature on batch correction, formally detecting batch heterogeneity (i.e., deciding whether correction is warranted) has received less attention. Existing approaches include variance-attribution heuristics such as PVCA \citep{li2009principal}, supervised PCA-based tests such as gPCA \citep{reese2013new}, related distributional tests (e.g., quantro \citep{hicks2015quantro}) and single-cell mixing scores (e.g., kBET  \citep{buttner2019test}) have also been proposed. However, in the clinical genomics regime, $p \gg n$ with small batches ($n \approx 20$--$30$),covariance-driven methods become unstable, distance-based methods suffer from high-dimensional concentration and dispersion confounding, and permutation tests provide coarse $p$-value resolution.

\section{Methods}\label{sec3}

Our principal focus is targeted amplicon sequencing for CNV detection. In this assay, predefined genomic loci (amplicons) are PCR-amplified and sequenced; the resulting read counts are approximately proportional to underlying copy number, with variability arising from PCR efficiency, sample effects, and sequencing noise.

Let $J$ denote the number of targeted amplicons retained after quality control for a given sample. For amplicon $j \in \{1,\ldots,J\}$, we observe a read count $s_j$ from the test sample and a corresponding count $n_j$ from a process-matched reference. In practice, $n_j$ may denote either (i) counts from a matched normal processed under comparable conditions,  (ii) a summary (e.g., median) across a set of normal samples processed together, or (iii) a summary across the entire batch (normal-free calling); we use the term reference to encompass both cases. We form log ratios
\begin{equation}
\tilde{y}_j = \log(s_j) - \log(n_j).
\end{equation}
Note that amplicons with low counts are filtered in QC, making zero-division errors not applicable. 

To remove global differences in sequencing depth and other sample-wide offsets, we center $\tilde{y}$ by subtracting a robust location estimate. Specifically, we set
\begin{equation}\label{eq:centering}
y_j = \tilde{y}_j - \operatorname{median}_{j' \in \mathcal{J}} \tilde{y}_{j'}
\quad \text{or} \quad
y_j = \tilde{y}_j - \operatorname{median}_{j' \in \mathcal{C}} \tilde{y}_{j'},
\end{equation}
where $\mathcal{J}$ is the set of all retained amplicons and $\mathcal{C} \subseteq \mathcal{J}$ is a predefined set of control amplicons known to be copy-neutral. After centering, $y_j \approx 0$ indicates copy neutrality, negative values indicate deletions, and positive values indicate amplifications. We refer to $\{y_j\}_{j=1}^J$ as log copy-number ratios (lCNRs).

\subsection{CNV Statistical Models and Evidence Computation}

Our heterogeneity-detection method requires a scalar evidence score for each sample that summarizes how compatible that sample's lCNR profile is with a CNV generative model. We employ two CNV models suited to different panel designs, enabling evaluation of evidence-based heterogeneity detection under distinct modeling assumptions. The first model treats amplicons within each gene as conditionally independent replicates and targets gene-level CNVs, appropriate for panels with sparse targets spread across the genome. The second model induces spatial dependence among neighboring amplicons via a state-space formulation, appropriate for densely tiled panels covering a contiguous region (in this work, BRCA1/2 and thalassemia assays). The two models also differ in inference procedures and in how evidence is estimated.

\subsubsection{Hierarchical Bayesian Model for Gene-Level CNV Detection}

In the gene-level model, amplicons targeting the same gene are treated as noisy measurements of an underlying gene-level lCNR. Let $g \in \{1,\ldots,G\}$ index genes, and let $y_{ig}$ denote the lCNR for the $i$th amplicon targeting gene $g$. The latent gene-level lCNR is $\mu_g$. We place hierarchical priors on gene-level means and gene-specific scales to share information across genes, and we use a SoftLaplace observation model, a smooth heavy-tailed alternative to the Laplace distribution, to improve robustness to outlier amplicons~\citep{ding2014three}.

The generative model is
\begin{equation}\label{eq:hierarchical}
\begin{aligned}
\mu_0 &\sim \mathcal{N}(0, 10), \\
\sigma^2 &\sim \mathrm{InvGamma}(\alpha_{\sigma}, \beta_{\sigma}), \\
\tau_0^2 &\sim \mathrm{InvGamma}(\alpha_{\tau_0}, \beta_{\tau_0}), \\
\mu_g \mid \mu_0, \sigma^2 &\sim \mathcal{N}(\mu_0, \sigma^2), \\
z_g &\sim \mathrm{InvGamma}(\alpha_{\tau}, \beta_{\tau}), \\
y_{ig} \mid \mu_g, z_g &\sim \mathrm{SoftLaplace}(\mu_g, \tau_0 z_g),
\end{aligned}
\end{equation}
where $\tau_0 z_g$ is a gene-specific scale parameter. The hyperparameters
$\{\alpha_{\sigma},\beta_{\sigma},\alpha_{\tau_0},\beta_{\tau_0},\alpha_{\tau},\beta_{\tau}\}$
are fixed a priori to values used in previous work.

We infer the model parameters
$\theta = \{\mu_0, \sigma^2, \tau_0^2, \{\mu_g\}_{g=1}^G, \{z_g\}_{g=1}^G\}$
using Hamiltonian Monte Carlo (HMC)~\citep{betancourt2017conceptual}. To estimate the log marginal likelihood $\log p(y)$, we use thermodynamic integration~\citep{calderhead2009estimating}. Let $\tau \in [0,1]$ denote the inverse temperature, and define the power posterior
\begin{equation}
p_{\tau}(\theta \mid y) \propto p(y \mid \theta)^{\tau}\,p(\theta).
\end{equation}
Thermodynamic integration uses the identity
\begin{equation}\label{eq:ti_property}
\ell = \log p(y) = \int_{0}^{1} \mathbb{E}_{\theta \sim p_{\tau}(\theta \mid y)} \big[\log p(y \mid \theta)\big] \, d\tau.
\end{equation}
We approximate the integral in~\eqref{eq:ti_property} by evaluating the expectation at a discrete ladder of inverse temperatures $\{\tau_t\}_{t=1}^{T}$ spanning $[0,1]$ and applying the trapezoidal rule. We use 100 temperatures with linear spacing.

\subsubsection{State-Space Model for Spatially Correlated CNV Detection}

When amplicons target a contiguous genomic region, lCNRs exhibit spatial correlation, and we instead use a state-space model~\citep{chopin2020introduction}. Let $x_j$ denote the latent true lCNR at amplicon $j$. Spatial dependence is induced via a heavy-tailed innovation model that favors local smoothness while permitting rare large transitions corresponding to CNV boundaries:
\begin{equation}\label{eq:transition_kernel}
p(x_{j+1} \mid x_j) = (1-p)\,\mathcal{N}(x_j, \epsilon) + p\,\mathcal{N}(x_j, \sigma^2),
\end{equation}
where $p \ll 1$ is the probability of a large transition, $\epsilon$ controls local smoothness, and $\sigma^2 \gg \epsilon$ permits large jumps. The initial state has a weakly informative prior reflecting the assumption of copy neutrality:
\begin{equation}
x_0 \sim \mathcal{N}(0, \tau_0^2).
\end{equation}
Observations follow a Laplace likelihood for robustness to outliers:
\begin{equation}
y_j \mid x_j \sim \mathrm{Laplace}(x_j, b).
\end{equation}

The complete model is
\begin{equation}\label{eq:state_space_model}
\begin{aligned}
x_0 &\sim \mathcal{N}(0, \tau_0^2), \\
x_{j+1} \mid x_j &\sim (1-p)\,\mathcal{N}(x_j, \epsilon) + p\,\mathcal{N}(x_j, \sigma^2), \\
y_j \mid x_j &\sim \mathrm{Laplace}(x_j, b),
\end{aligned}
\end{equation}
with fixed transition and observation parameters
$\theta = \{p, \epsilon, \sigma^2, \tau_0^2, b\}$ chosen a priori for each assay using domain knowledge of expected CNV sizes and noise characteristics~\citep{talbot2025classifying}. Full Bayesian inference over $\theta$ is computationally prohibitive for our use case, and we therefore treat $\theta$ as fixed when computing evidence.

Inference over latent states $\{x_j\}_{j=0}^{J}$ proceeds via sequential Monte Carlo using an auxiliary particle filter~\citep{pitt1999filtering}. A standard byproduct of particle filtering is an unbiased estimator of the marginal likelihood. Let $M$ denote the number of particles and let $w_j^{(m)}$ denote the (unnormalized) importance weight for particle $m$ at position $j$. The particle filter yields the estimator
\begin{equation}\label{eq:pf_evidence}
\ell =  \log \widehat{p}(y \mid \theta) = \sum_{j=1}^{J}\log\left(\frac{1}{M}\sum_{m=1}^{M} w_j^{(m)}\right).
\end{equation}

\begin{figure}[ht]
\centering
\includegraphics[width=1.0\textwidth]{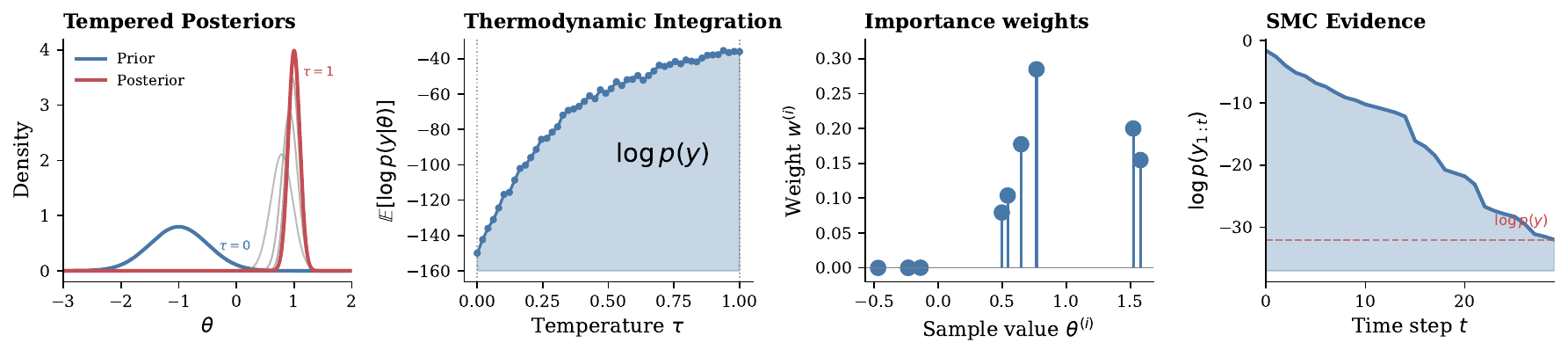}
\caption{
Overview of evidence computation. Left: a sequence of power posteriors indexed by inverse temperature $\tau \in [0,1]$ used for thermodynamic integration. Second from left: thermodynamic integration estimates $\log p(y)$ by numerically integrating $\mathbb{E}_{p_{\tau}(\theta\mid y)}[\log p(y\mid \theta)]$ over $\tau$ (Eq.~\eqref{eq:ti_property}). Middle right: particle-filter weights at a representative amplicon position; highly peaked weight distributions indicate that few particles explain the observation well. Right: particle-filter evidence accumulates across amplicons via the product form in Eq.~\eqref{eq:pf_evidence}.
}\label{fig:evidence}
\end{figure}

\subsection{Detecting Batch Heterogeneity via Mixture Modeling of Model Evidence}

We compute a normalized evidence score $\ell_i$ for each sample $i \in \{1,\ldots,N\}$ in a cohort. Our core hypothesis is that batch heterogeneity manifests as systematic differences in model compatibility across samples: if a cohort contains two subsets processed under different technical conditions, one subset may be consistently better (or worse) explained by the assumed CNV model, producing a bimodal (or otherwise mixture-like) distribution of $\{\ell_i\}$.

We formalize heterogeneity detection as a likelihood-ratio test comparing a one-component model (homogeneous cohort) to a two-component mixture (heterogeneous cohort). Specifically, we consider
\begin{align}
\text{Null (H$_0$)}: \quad \ell_i &\sim f(\ell \mid \mu_1, \alpha_1), \\
\text{Alternative (H$_1$)}: \quad \ell_i &\sim \pi\, f(\ell \mid \mu_1, \alpha_1) + (1 - \pi)\, f(\ell \mid \mu_2, \alpha_2),
\end{align}
where $\pi \in (0,1)$ is the mixture weight and $f(\cdot \mid \mu,\alpha)$ is a location-scale density. We consider several choices of $f$ which we consider in this work, namely Gaussian, Laplace, hypersecant, t-distribution, and gennorm. We estimate parameters under $H_0$ and $H_1$ via expectation-maximization (EM). For the two-component mixture, we initialize EM using $k$-means centers and perform $n_{\text{init}}=3$ random restarts, retaining the solution with the highest final log likelihood. Convergence is declared when the change in average log likelihood is below $10^{-4}$ or after 1000 iterations.

\begin{algorithm}[ht]
\caption{Parametric Bootstrap Likelihood Ratio Test in Evidence Space}
\label{alg:bootstrap_lrt}
\KwIn{Evidence scores $\{\ell_i\}_{i=1}^N$, number of bootstrap replicates $B$}
\KwOut{$p$-value and likelihood ratio statistic $\text{LR}_{\text{obs}}$}

\BlankLine
Fit 1-component generalized normal model $\to \widehat{M}_0$\;

Fit 2-component generalized normal mixture $\to \widehat{M}_1$\;

$\text{LR}_{\text{obs}} \gets 2 \cdot (\mathcal{L}(\{\ell_i\} \mid \widehat{M}_1) - \mathcal{L}(\{\ell_i\} \mid \widehat{M}_0))$\;

\BlankLine
\For{$b \gets 1$ \KwTo $B$}{
    Sample $\{\ell_i^{(b)}\}_{i=1}^N \sim \widehat{M}_0$\;

    Fit 1-component model to $\{\ell_i^{(b)}\} \to \widehat{M}_0^{(b)}$\;

    Fit 2-component mixture to $\{\ell_i^{(b)}\} \to \widehat{M}_1^{(b)}$\;

    $\text{LR}_b \gets 2 \cdot (\mathcal{L}(\{\ell_i^{(b)}\} \mid \widehat{M}_1^{(b)}) - \mathcal{L}(\{\ell_i^{(b)}\} \mid \widehat{M}_0^{(b)}))$\;
}
\BlankLine
$p \gets \frac{1 + \sum_{b=1}^B \mathbb{I}(\text{LR}_b \geq \text{LR}_{\text{obs}})}{B + 1}$\;

\Return{$p$, $\text{LR}_{\text{obs}}$}
\end{algorithm}

Because finite-mixture models are non-regular (the null hypothesis lies on the boundary of the alternative parameter space), the asymptotic chi-square approximation for the likelihood ratio statistic does not apply~\citep{mclachlan2004finite}. We therefore estimate the null distribution of the likelihood ratio statistic via a parametric bootstrap~\citep{davison1997bootstrap}.

\begin{figure}[ht]
\centering
\includegraphics[width=1.0\textwidth]{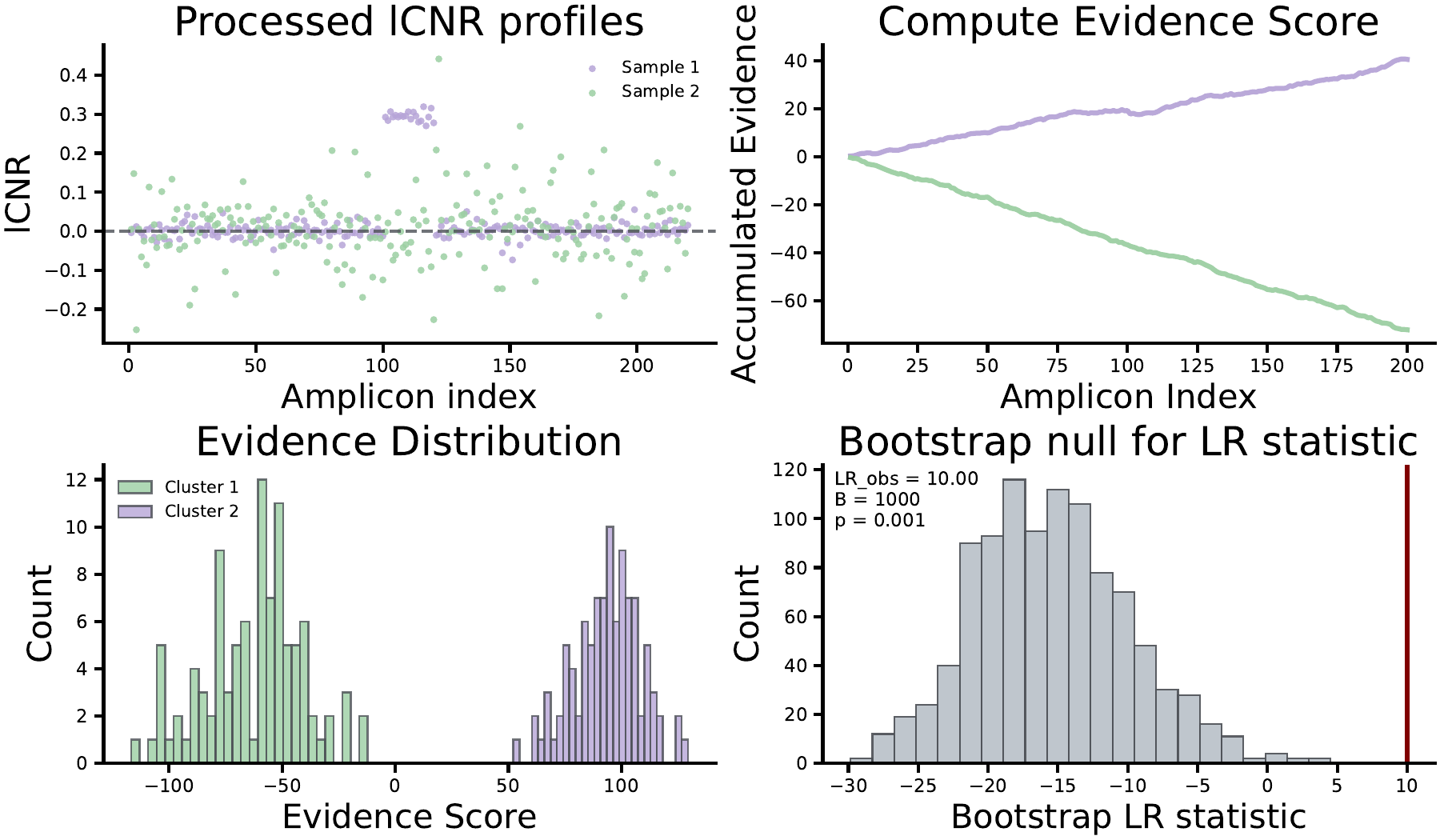}
\caption{
Overview of the proposed heterogeneity detection workflow. On the top left we show two sample LCNrs after preprocessing according to~\eqref{eq:centering}. The top right shows an example computation of the evidence, matching the right of Figure~\ref{fig:evidence}. Bottom left shows the distribution of the evidences of the samples in the entire batch with a clear bimodal structure and color corresponding to group label. Finally, bottom right visualizes the significance test defined by Algorithm~\ref{alg:bootstrap_lrt}.
}\label{fig:overview}
\end{figure}

Let $\widehat{M}_0$ and $\widehat{M}_1$ denote the fitted one- and two-component models, respectively, and let $\mathcal{L}(\cdot \mid M)$ denote the log likelihood under model $M$. We compute the observed likelihood ratio statistic
\begin{equation}
\text{LR}_\text{obs} = 2\Bigl(\mathcal{L}(\{\ell_i\} \mid \widehat{M}_1) - \mathcal{L}(\{\ell_i\} \mid \widehat{M}_0)\Bigr).
\end{equation}
For bootstrap replicate $b=1,\ldots,B$ (we use $B=1000$), we draw an i.i.d.\ sample
\begin{equation}
\{\ell_i^{(b)}\}_{i=1}^{N} \sim \widehat{M}_0,
\end{equation}
refit both one- and two-component models to $\{\ell_i^{(b)}\}$, and compute
\begin{equation}
\text{LR}_b = 2\Bigl(\mathcal{L}(\{\ell_i^{(b)}\} \mid \widehat{M}_1^{(b)}) - \mathcal{L}(\{\ell_i^{(b)}\} \mid \widehat{M}_0^{(b)})\Bigr).
\end{equation}
We then estimate a finite-sample $p$-value as
\begin{equation}
p = \frac{1 + \sum_{b=1}^{B} 1_{\text{LR}_b \geq \text{LR}_\text{obs}}}{B + 1},
\end{equation}
which includes the observed statistic itself~\citep{davison1997bootstrap}. Small $p$-values indicate that the evidence-score distribution is unlikely to arise from a single homogeneous population, suggesting batch heterogeneity.

\section{Synthetic Results: A Conservative Null Distribution}\label{sec4}
We first assess calibration on synthetic data under a null in which technical batch effects are absent, but biological signal may be present. Specifically, we generate lCNR data from the hierarchical mechanism in \eqref{eq:hierarchical} using settings matched to a targeted liquid-biopsy panel containing 439 amplicons, including a subset targeting 10 clinically relevant genes. We simulate 1000 independent batches, each comprising $N=20$ samples. For each simulated batch, we construct a synthetic reference by taking the per-amplicon mean across the batch (the ``batch mean'' reference), fit the hierarchical CNV model, compute per-sample evidence scores, and apply our heterogeneity detection procedure to obtain a batch-level $p$-value. We repeat this analysis while varying the component density used in the evidence-space mixture model, comparing five candidates: Gaussian, Student-$t$ with $\nu=3$, generalized normal (gennorm), Laplace, and hypersecant (sech). This experiment isolates the sensitivity of the downstream mixture test to the choice of evidence-space distribution, holding the data-generating mechanism and evidence computation fixed. We perform a second experiment where we artificially subtract two standard deviations of the ``worst'' sample, artificially introducing an outlier into the data.

Accurately diagnosing $p$-value calibration can require many more null replicates than are practical here~\citep{storey2003statistical}, because each replicate entails evidence computation for 20 samples and thus thermodynamic integration for 20{,}000 synthetic cases overall even for the 1000 batches considered here. We therefore adopt two pragmatic criteria that directly target conservativeness at clinically relevant thresholds.  First, under a well-calibrated null, $p$-values are uniformly distributed, i.e., $\mathrm{Beta}(1,1)$. To summarize departures from uniformity in a parsimonious and interpretable way, we model the empirical $p$-value distribution as
\begin{equation}
p_i \sim  \mathrm{Beta}(a,1), \qquad i=1,\ldots,n,
\end{equation}
where $a>1$ indicates conservativeness (density is increasing in $p$, so small $p$-values are depleted), $a\approx 1$ indicates approximate calibration, and $a<1$ indicates anti-conservativeness (small $p$-values are enriched).
With a Gamma prior $a \sim \mathrm{Gamma}(\alpha,\lambda_0)$ (shape $\alpha$, rate $\lambda_0$), the posterior is conjugate. We use the 95\% credible interval to quantify uncertainty in $a$. Second, we place an upper bound on the probability of rejection under the null at $0.05$. 
Let $X = \sum_{i=1}^n 1_{p_i < 0.05}$ denote the number of null batches that would be flagged at a nominal 0.05 threshold. Treating $X \sim \mathrm{Binomial}(n,q)$ with $q=\mathbb{P}(p<0.05)$ under the null, we compute a Clopper--Pearson upper confidence bound for $q$~\citep{lehmann2005testing}. This provides a direct, clinically interpretable bound on the false-alarm probability at the operating point used in practice.

Table~\ref{tab:synthetic} summarizes both criteria across evidence-space component distributions. We can see that except for the Laplace distribution, all distributions have comparable properties of a nearly uniform distribution under the null, with alpha estimates ranging from $0.9$ to $1.01$ and the Clopper---Pearson upper interval ranging from $0.025$ to $0.07$. When we introduced an outlier, however, the situation changed dramatically. Naturally, the Gaussian distribution became highly anti-conservative with $\alpha=0.25 (0.23,0.27)$, with gennorm and student-t affected to a lesser extent. Surprisingly, the Laplace distribution was the best calibrated of the mixture likelihoods. Figure~\ref{fig:synth_pdist} in the Appendix visualizes the resulting distribution.

For completeness, Appendix~\ref{app:additional_results} compares our parametric-bootstrap test to a non-bootstrapped alternative, the Lo--Mendell--Rubin (LMR) test~\citep{lo2001testing} and Bayesian model selection. We can see in Table~\ref{tab:synthetic_lmr} that here the component distribution makes a substantial difference, with the hypersecant distribution overly conservative, while the Gaussian and Laplace distributions are highly anti-conservative. As such, the bootstrap procedure is important in clinical applications.

\begin{table}[ht]
\centering
\begin{tabular}{l|cc|cc}
 & \multicolumn{2}{c|}{\textbf{No Outlier}} 
 & \multicolumn{2}{c}{\textbf{Outlier Included}} 
 \\
\textbf{Distribution} 
 & \textbf{Alpha estimate} & \textbf{Clopper-Pearson}
& \textbf{Alpha estimate} & \textbf{Clopper-Pearson}
 \\
\midrule
Gaussian & $0.96\;(0.90,1.02)$ & $0.07$ & $0.25\;(0.23,0.27)$ & $0.73$ \\
Hypersecant & $0.915\;(0.86,0.97)$ & $0.06$ & $1.53\;(1.43,1.62)$ & $0.06$ \\
Student-t & $1.01\;(0.96,1.08)$ & $0.025$  & $0.62\;(0.58,0.66)$ & $0.32$ \\
Laplace & $0.7561\;(0.71, 0.8)$ & $0.12$ & $0.94 (0.89, 1.00)$ & $0.06$ \\
Gennorm & $0.93\;(0.88,0.99)$ & $0.05$  & $0.44\;(0.41,0.47)$ & $0.41$ \\
\midrule
\end{tabular}
\caption{Effect of mixture distribution on conservativeness}
\label{tab:synthetic}
\end{table}

\section{Cohort} \label{sec3p2}
% Requires (at minimum):
% \usepackage{booktabs}
% \usepackage{natbib} % if using \citep/\citet

A detailed description of preprocessing and quality control is provided in
Appendix~\ref{dataprocessing}. Here we summarize the datasets used for evaluation,
emphasizing sample sizes, batch factors, and biological signals relevant to
batch-effect detection. We consider two modalities: targeted amplicon sequencing
(three clinical panels, $n=24$--$100$) and mouse
electrophysiology (a combination of two studies with $n=54$). Key
dataset characteristics are summarized in Table~\ref{tab:datasets}.

\begin{table}[t]
\centering
\caption{Summary of datasets used for batch effect algorithm validation.}
\label{tab:datasets}
\begin{tabular}{lcccl}
\toprule
\textbf{Dataset} & \textbf{N} & \textbf{Batch Factor} & \textbf{Bio.\ Signal} & \textbf{Validation Purpose} \\
\midrule
Liquid Biopsy      & 32  & FFPE quality & None     & Technical artifact robustness \\
BRCA Panel         & 24  & Reagent lot  & CNV+     & Batch vs.\ biology discrimination \\
Thalassemia Panel  & 100 & Laboratory   & CNV+     & Multi-site heterogeneity \\
Electrophysiology  & 54  & Experiment   & Behavior & Cross-modality generalization \\
\bottomrule
\end{tabular}
\end{table}

\subsection{Targeted Amplicon Sequencing Data}
We evaluate three targeted panels spanning distinct, practically relevant
batch-effect mechanisms; panel design details and preprocessing are provided in
Appendix~\ref{dataprocessing}.

\subsubsection{Multi-cancer Liquid Biopsy Panel (LBX)}
This dataset comprises 32 FFPE samples sequenced with a 173-amplicon panel
designed for low-cost cancer screening (targeting five CNV-associated genes).
The batch factor is FFPE preservation quality: 10 libraries were prepared from
substantially fragmented FFPE DNA, and 22 from higher-quality clinical FFPE
material. This cohort evaluates robustness to amplification-related technical
heterogeneity induced by mixed sample quality.

\subsubsection{BRCA Panel}
The BRCA dataset targets \textit{BRCA1} and \textit{BRCA2} using 283 amplicons
(including normalization controls). The batch factor is reagent lot: 24 samples
were sequenced across two lots (lot~A: $n=16$; lot~B: $n=8$), introducing
systematic shifts in amplification efficiency and baseline coverage profiles.
Biological signal is present via CNVs in 12 samples (six with gene-level
\textit{BRCA2} gains/losses; six with exon-level \textit{BRCA1} aberrations),
enabling assessment of whether heterogeneity is correctly attributed to batch
effects rather than true copy-number variation.

\subsubsection{Thalassemia Panel}
The thalassemia dataset uses a 120-amplicon panel (including normalization
controls) targeting the alpha- and beta-globin loci. The batch factor is
laboratory origin in a two-site setting (Laboratory~A: $n=58$; Laboratory~B:
$n=42$), reflecting multi-site heterogeneity from differences in protocols,
reagents, instrumentation, and operator effects. Biological heterogeneity is
substantial: 55 samples (55\%) carry CNVs affecting the alpha-globin locus, and
four additional samples are known poorly-modeled outliers reported
in~\citet{talbot2025classifying}; we retain these outliers to evaluate behavior
under non-ideal model fit. Overall, this cohort evaluates batch-effect detection
in a realistic clinical genetics scenario combining multi-site technical
variation with strong biological signal.

\subsection{Mouse Electrophysiology Data}
To assess cross-modality generalization beyond genomics, we include local field
potential (LFP) recordings from 54 mice aggregated from two experiments that
serve as the batch factor. The first study (SP) recorded LFPs from 26 mice
across 8 brain regions during social (paired interaction) and non-social
(isolated exploration) conditions~\citep{mague2022brain}. The second study (TST) recorded from 28 mice across 11
brain regions during stress-exposure and control conditions; 7 brain regions overlapped with SP \citep{talbot2023estimating}.
Feature processing is described in Appendix~\ref{dataprocessing}. Within each
experiment, protocols were highly controlled (minimizing intra-experiment batch
effects), however, the data in SP was collected in two sequential cohorts. As such, we expect no batch effects in the TST experiment, possible effects in the SP experiment, and likely effects when the data from the two experiments are considered jointly. This dataset
therefore evaluates algorithm performance in a setting where batch effects are
not reliably diagnosable by visual inspection and where domain shift is expected
across experiments.

\section{Results} \label{sec5}
\subsection{CNV Heterogeneity Detection}

\subsubsection{Overall Performance}

In our first analysis, we demonstrate the performance of our method in detecting batch effects in each of the three datasets. We used each of the kernels to evaluate sensitivity to the choice of mixture model, and used 10000 bootstrap iterations for computing the p-values. We also evaluated these p-values on each of the within-batch groups to evaluate the predictions when no batch effects are expected. The results are shown in Table~\ref{tab:pvalues}. 

For the thalassemia and LBX groups, batch effects were significant with every mixture model except hypersecant (thalassemia $p=0.16$). This conservativeness is to be expected based on our results in Section~\ref{sec4}. The p-values were particularly small in thalassemia, which is to be expected due to its relatively large sample size. With the BRCA panel, the results were noticeably weaker, with only the Laplace mixture yielding a significant result. Presumably, the large CNVs combined reduced the ability to distinguish the two groups. 

\begin{table}[ht]
\centering
\begin{tabular}{l|cc|cc|cc}
 & \multicolumn{2}{c|}{\textbf{BRCA}} 
 & \multicolumn{2}{c|}{\textbf{Thalassemia}} 
 & \multicolumn{2}{c}{\textbf{LBX FFPE}} \\
\textbf{Distribution} 
 & \textbf{Overall} & \textbf{Subgroups}
 & \textbf{Overall} & \textbf{Subgroups}
 & \textbf{Overall} & \textbf{Subgroups} \\
\midrule
Gaussian & $0.070$ & $0.130,0.477$ & $0.001$ & $0.023,0.032$ & $0.013$ & $0.352,0.499$ \\
Hypersecant & $0.800$ & $0.177,0.524$ & $0.161$ & $0.069,0.155$ & $0.007$ & $0.682,0.108$ \\
Student-t & $0.077$ & $0.122,0.477$ & $0.005$ & $0.059,0.180$ & $0.031$ & $0.303,0.294$ \\
Laplace & $0.028$ & $0.057,0.110$ & $0.001$ & $0.118,0.650$& $0.011$ & $0.233,0.141$ \\
Gennorm & $0.055$ & $0.092,0.743$ & $0.001$ & $0.039,0.101$ & $0.016$ & $0.261,0.264$ \\
\end{tabular}
\caption{P-values of the three CNV panels, both the aggregate data with batch effects (Overall), and the two subgroups presumably lacking batch effects (Subgroups).}
\label{tab:pvalues}
\end{table}

Notably, however, none of the subgroups with each of these experiments yielded a significant p-value with the lowest p-value for BRCA and the LBX panels being $0.057$ and $0.108$ respectively. One subgroup in the thalassemia group was borderline significant ($p=0.04$ for gennorm), but nowhere near the significance level seen in the combined group. This also matches our conclusions on our tests being conservative, even though each group in the thalassemia and BRCA panels have a mixture of positive and negative samples, none are flagged as having batch effects.

\subsubsection{Statistical Power and Size}

\begin{figure}[ht]
\centering
\includegraphics[width=1.0\textwidth]{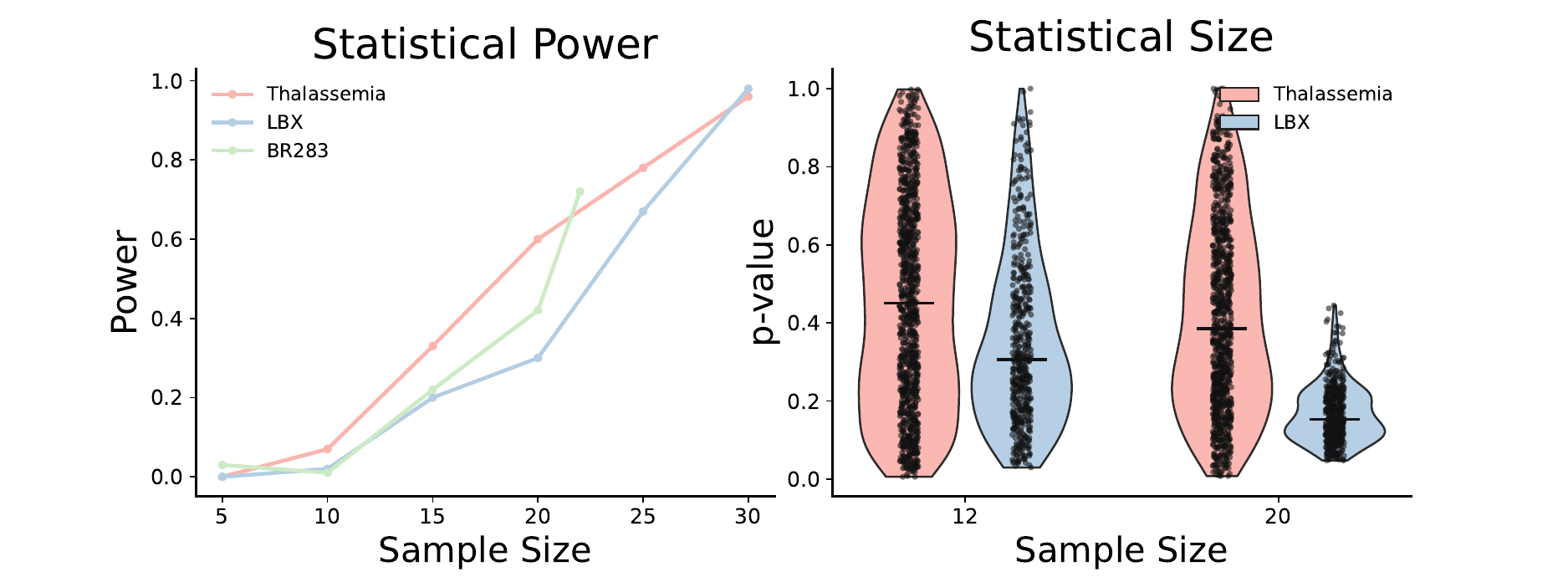}
\caption{
The left figure shows statistical power as a function of sample size in each of the three CNV assays. The right figure shows the distributions of p-values at given sample sizes for the thalassemia and LBX assays.
}\label{fig:power_size}
\end{figure}

We then evaluate the statistical power and size of our approach in detecting batch effects on each of these panels as a function of sample sizes. We evaluate the former by repeatedly resampling the data from both groups from sample  sizes ranging from 5 to 30, covering the batch sizes we commonly encounter from our customers. To evaluate size, we fit a model with samples coming exclusively from a single group at sample sizes of 12 and 20. The results are shown in Figure~\ref{fig:power_size}.

We can see that statistical power is low in all experiments at sample sizes less than 15. However, by 25 and 30 samples power was greater than $0.8$. Given that most introductory tests have comparable sample sizes, this is sufficient to determine if the process is sufficiently matched for calling. We also found statistical size was acceptable for thalassemia and LBX at both sample sizes, with a maximum of 4.4\% of values being less than $0.05$. A plot of the distribution of these p-values is given on the right of Figure~\ref{fig:power_size}. The LBX distribution is artificially compressed, likely due to resampling artifacts (selecting a sample size of 20 from 24 samples introduces substantial redundancy). 

\subsubsection{Clustering Accuracy}
\label{ssec:accuracy}

We then evaluate the ability of our method to select comparable samples, allowing us to directly compare to commonly-used methods in the field. We compare our approach against two families of unsupervised advocated for in~\citet{kusmirek2019comparison}. The first family of methods 
operates directly on self-normalized read count matrices. Given a data matrix 
$\mathbf{X} \in \mathbb{R}^{n \times p}$ with $n$ samples and $p$ features 
(e.g., amplicons), we normalize each sample by dividing by its mean count 
(with a pseudocount of 1 to avoid division by zero). Clustering is then 
performed either via Gaussian mixture models (GMM), which assume samples arise 
from a mixture of multivariate Gaussian distributions, or via dimensionality 
reduction followed by hierarchical clustering. For the latter, we apply 
principal component analysis (PCA) to the normalized counts, retain the top 
$k$ components, and cluster the resulting embeddings using Ward's 
minimum-variance linkage. We refer to these methods as \textbf{GMM} and 
\textbf{PCA-Ward}, respectively.

The second family of methods first 
computes pairwise sample correlations from the self-normalized counts, yielding 
a symmetric $n \times n$ correlation matrix $\mathbf{R}$. We evaluate both 
Pearson and Spearman correlations, with diagonal elements set to zero to 
exclude self-similarity. Clusters are then identified using one of three 
approaches: (1)~\textbf{Hierarchical}, which converts correlations 
to distances ($d_{ij} = 1 - r_{ij}$) and applies agglomerative clustering with 
average linkage; (2)~\textbf{Spectral}, which constructs an affinity 
matrix $\mathbf{S}$ with $S_{ij} = \exp(\gamma \cdot \max(r_{ij}, 0))$ and 
applies spectral embedding followed by $k$-means; and (3)~\textbf{PCA-KMeans}, 
which performs eigendecomposition of the correlation matrix, retains the top 
eigenvectors as a low-dimensional sample embedding, and applies $k$-means 
clustering. All methods require specifying the number of clusters $k$ a priori, which we set to the known value of 2.  

\begin{table}[ht]
\centering
\begin{tabular}{l|cc|cc|cc}
 & \multicolumn{2}{c|}{\textbf{BRCA}} 
 & \multicolumn{2}{c|}{\textbf{Thalassemia}} 
 & \multicolumn{2}{c}{\textbf{LBX FFPE}} \\
\textbf{Model} 
 & \textbf{Accuracy} & \textbf{Kappa}
 & \textbf{Accuracy} & \textbf{Kappa}
 & \textbf{Accuracy} & \textbf{Kappa} \\
\midrule
GMM & 0.63 & 0.0 & 0.78 & 0.55 & 0.91 & 0.76 \\
PCA + Ward & 0.63 & 0.0 & 0.76 & 0.51 & 0.91 & 0.76 \\
\midrule
Hierarchical & 0.63 & 0.0 & 0.78 & 0.55 & 0.72 & 0.13 \\
Spectral & \textbf{0.71} & 0.27 & 0.84 & 0.66 & 0.94 & 0.85 \\
PCA-KMeans & 0.63 & 0.0 & 0.85 & 0.68 & 0.94 & 0.86 \\
\midrule
Likelihood (Ours) & 0.63 & \textbf{0.34} & \textbf{0.90} & \textbf{0.80} & \textbf{1.0} & \textbf{1.0} \\
\midrule
\end{tabular}
\caption{Batch selection performance of our method to competitors in the BRCA, Thalassemia, and LBX assays.}
\label{tab:multiassay_comparison}
\end{table}

Our likelihood-based approach achieves the best or tied-best performance across 
all three targeted sequencing datasets (Table~\ref{tab:multiassay_comparison}). 
On the LBX FFPE dataset, our method achieves perfect batch classification 
(accuracy = 1.0, $\kappa$ = 1.0), compared to the next-best methods (Spectral 
and PCA-KMeans) which achieve 0.94 accuracy and $\kappa \leq 0.86$. On the 
Thalassemia dataset, our method attains 0.90 accuracy and $\kappa = 0.80$, 
representing a 5--14 percentage point improvement in accuracy and a 12--29 
point improvement in Cohen's $\kappa$ over all baselines. These gains are 
particularly notable given that the Thalassemia dataset contains 55\% 
CNV-positive samples, demonstrating that our approach can reliably distinguish 
laboratory-driven batch effects from biological heterogeneity even when the 
latter is substantial.

The BRCA dataset presents a uniquely challenging scenario: 12 of 24 samples 
harbor germline gene-level CNVs in \textit{BRCA2}, which comprises the majority 
of amplicons in the panel. This biological signal induces strong 
within-phenotype similarity that cuts across reagent lots, causing most methods 
to predict the majority class and achieve $\kappa = 0$. While Spectral 
clustering attains the highest raw accuracy (0.71), it does so with only 
moderate agreement ($\kappa = 0.27$). In contrast, our likelihood-based method 
achieves the highest $\kappa$ (0.34) despite matching the baseline accuracy, 
indicating that it captures meaningful batch structure rather than simply 
exploiting class imbalance. This result underscores a key advantage of our 
approach: by explicitly modeling the generative process underlying read counts, 
it can disentangle technical batch effects from confounding biological signal 
where correlation- and distance-based methods fail.

\subsection{Mouse LFP Data}

Finally, we evaluate our method on two electrophysiology datasets of psychiatric disorders in mice. We use probabilistic principal component analysis (PPCA) as our generative model of choice, defined as
\begin{equation}
    \begin{split}
        p(z)&=N(0,I_L)\\
        p(x|z,W,\sigma^2)&=N(Wz,\sigma^2I)
    \end{split}
\end{equation}
Unfortunately, the temporal correlation between the observations violate the assumptions of almost all dimensionality-selection techniques, so 20 components were chosen to roughly match the dimensionality used in the original works.  This model was fit on a training set of 10 mice from the TST dataset and 5 from the SP dataset. We then evaluated the log likelihood of each observation in the holdout datasets. We then averaged the likelihoods within each mouse within condition, resulting in two observations per mouse. This step was done due to the incredible noise in the observations (the spectral features are incredibly noisy), particularly at low frequencies.

Table~\ref{tab:electrophysiology_comparison} summarizes the results.  Unsurprisingly, there was incredibly strong evidence of batch heterogeneity when including mice from both experiments, with an average p-value of $0.005$. What is interesting is that the TST dataset did not detect batch effects, while SP did. Tail suspension is incredibly stressful and trivial to detect, but our algorithm was not triggered even when the positive and negative conditions were separated. On the other hand, the SP experiment indicated the presence of batch effects ($p=.014$). These mice were recorded in two batches, an initial cohort was augmented by a second cohort several years later. This distinction is maintained, when conditions are separated, but weaker. A potential explanation for this behavior is that the two conditions are nearly indistinguishable, so separating the two conditions increases the variance/sample size without introducing an additional cluster.

\begin{table}[ht]
\centering
\begin{tabular}{l|cc|cc}
 & \multicolumn{2}{c|}{\textbf{Conditions Averaged}} 
 & \multicolumn{2}{c}{\textbf{Conditions Separated}} 
 \\
\textbf{Experiment} 
 & \textbf{Significance Prob} & \textbf{Avg P-Value}
 & \textbf{Significance Prob} & \textbf{Avg P-Value}
 \\
\midrule
Social Preference & 1.0 & 0.026 & 1.0 & 0.010 \\
Tail Suspension & 0.0 & 0.67 & 0.0 & 0.12 \\
Both & 1.0 & 0.011 & 1.0 & 0.004  \\
\midrule
\end{tabular}
\caption{Batch effect detection in the electrophysiology data}
\label{tab:electrophysiology_comparison}
\end{table}

\section{Discussion} \label{sec6}
Batch effects are a serious concern in many clinical and scientific applications. These issues are particularly acute in CNV detection algorithms requiring normal samples to adjust for amplicon-specific effects. As such, a reliable method for detecting such effects, and potentially selecting process-matched samples is critical. In this work, we develop a novel method for performing this task based on the Bayesian model evidence. This quantity measures how informative a sample is, or equivalently how well the model describes the data, and as such provides a natural metric for detecting a lack of process matching. We show that our method has many desirable properties; it is robust, statistically conservative, and broadly applicable.

\bibliography{sample}

\newpage
\appendix
\section{Data Processing}
\label{dataprocessing}

\subsection*{Targeted Amplicon Sequencing}

We first performed read-to-genome alignment to the GRCh37/hg19 genome using BWA-MEM \citep{li2013aligning}. The initial alignment is improved by performing local realignment using Smith-Waterman \citep{smith1981identification} and a proprietary algorithm. To maximize the base accuracy and minimize the sequencing noise, paired-end reads are assembled into consensus reads, weighted with the base quality scores from both mates. The assembled reads correspond to the gene-specific positions on the genome. A set of filters are applied to remove non-uniquely mapped reads (e.g., pseudogenes) and reads that do not match the amplicon positions (e.g., primer-dimers or non-specific amplifications). For CNV calculations, per-amplicon read depth is calculated for each sample. Counts of overlapping amplicons are stored as independent entries.

\subsubsection*{Liquid Biopsy Data}

Our first dataset came from a liquid biopsy panel designed to cover common carcinogenic mutations. It had 173 amplicons targeting a variety of genes, 5 of which were CNV targets.

\subsubsection*{Thalassemia Data}

For the second group, we obtained 44 positive and 13 negative clinical samples as de-identified remnants from a laboratory that tested the samples for clinical use and would have otherwise discarded them. FDA guidance states that such samples may be used for research purposes without obtaining patient consent. These samples were sequenced using a panel of 131 amplicons (of which 11 are excluded gap amplicons) that covers alpha, beta, delta, gamma and epsilon Thalassemia regions, as well as pseudogenes and control regions. Sequencing was performed on an Illumina MiSeq$^{\mathrm{TM}}$ platform to an average depth of approximately 2800 paired-end reads. The set of true alpha thalassemia diagnoses were obtained from an orthogonal, clinically validated test that uses MLPA with reflex to GAP-PCR for alpha-thalassemia deletion and duplication detection. Of the positive samples, 31 had a heterozygous alpha-3.7 deletion, 5 had a homozygous alpha-3.7 deletion, 2 had alpha-4.2 deletions, 3 had south-east Asian deletions, 1 sample with an alpha-4.2 duplication, 1 sample with a large HBA duplication, and 1 sample with an alpha-3.7 deletion/4.2 duplication. The second group of samples was obtained similarly from a separate laboratory, and contained 43 samples.

\subsubsection*{BRCA Data}

To further investigate our quality control metrics, we obtained a second dataset sequenced using a panel designed to detect mutations in the BRCA1 and BRCA2 genes. This panel contains 283 amplicons targeting various regions within BRCA1 and BRCA2 genes, as well as 8 control amplicons from other chromosomes. We obtained four mutation-positive DNA cell line samples from the Coriell Institute for Medical Research with known mutations: NA18949 with a BRCA1 exon 14 and 15 deletion, NA14626 with a BRCA1 exon 12 duplication, NA0330 with a whole-gene duplication of BRCA2, and NA02718 with a whole gene deletion of BRCA2. We also obtained four negative cell line samples NA12878, NA19240, NA24385, and NA24143. Additionally, Horizon Discovery's Mimix$^{\mathrm{TM}}$ Quantitative Multiplex, fcDNA (Moderate) Reference Standard was used to evalue performance on Formalin-compromised DNA (FFPE). Libraries were prepared and sequenced on Illumina's MiSeq$^{\mathrm{TM}}$. All samples were collected under informed consent. 

\subsection*{Mouse Electrophysiology}

\subsubsection*{Tail Suspension Test}
The LFPs analyzed in the Tail Suspension Test (TST) came from 26 mice of two genetic backgrounds (14 wild type and 12 Clock-$\Delta$19). The Clock-$\Delta$19 genotype has been used to model bipolar disorder \citep{VanEnkhuizen2013a}. Each mouse was recorded for 20 minutes across 3 behavioral contexts: 5 minutes in its home cage (non-stressful), 5 minutes in an open field (arousing, mildly stressful), and 10 minutes suspended by its tail (highly stressful). Data were recorded from 11 biologically relevant brain regions with 32 electrodes (multiple electrodes per region) at 1000 Hz. These redundant electrodes were implanted to allow for the removal of faulty electrodes and electrodes that missed the target brain region. We chose to average the signals per region to yield an 11-dimensional time series per mouse. This was done because the brain region represents the smallest resolvable location when modeling multiple animals; multiple electrodes function as repeated measurements and averaging allows us to reduce the variance of the measured signal. 

We discretized the time series into one-second windows to model how these spectral characteristics change over time. Windows containing saturated signals were removed (typically due to motion artifacts). While there are methods that can characterize the spectral features on a continuous time scale \citep{prado2010time}, the behavior we deal with changes over a longer scale than the observed dynamics. Consequently, it is more effective to discretize and obtain sharper spectral feature estimates \citep{Cohen1995TimeFrequency} that are more amenable to factor modeling.

We chose to extract the relevant quantities from the recorded data prior to modeling rather than extracting spectral features in the modeling framework for simplicity; the extra modeling step would substantially increase the number of parameters in the model. The features related to power were computed from 1 Hz to 56 Hz in 1 Hz bands using Welch’s method \citep{welch1967use}, which is recommended in the neuroscience literature \citep{kass2014analysis}. We chose 56 Hz as a threshold to avoid the substantial noise induced at 60 Hz from the recording equipment, as prior literature has demonstrated that much of the meaningful information is contained in the lower frequencies.

\subsubsection*{Social Preference Dataset}

The social preference dataset used identical processing steps as the TST dataset. In this experiment, the mice were placed in a two-chamber assay where the mouse was allowed to wander freely with simultaneous recordings and position tracking for 10 minutes. One chamber included another mouse (social interaction), while the second chamber included an inanimate object (non-social interaction). The original purpose of the study was to characterize the electrophysiology associated with social behavior and causally manipulate these dynamics using optogenetic stimulation. A full description of the methods is given in~\citet{mague2022brain}. 

\subsubsection*{Data Availability}

All data from the targeted panels is located on the Github repository associated with this work. The TST dataset can be downloaded at \url{https://research.repository.duke.edu/concern/datasets/zc77sr31x?locale=en}. For access to the social preference dataset contact the principal investigator, Kafui Dzirasa.

\begin{figure}[ht]
\centering
\includegraphics[width=1.0\textwidth]{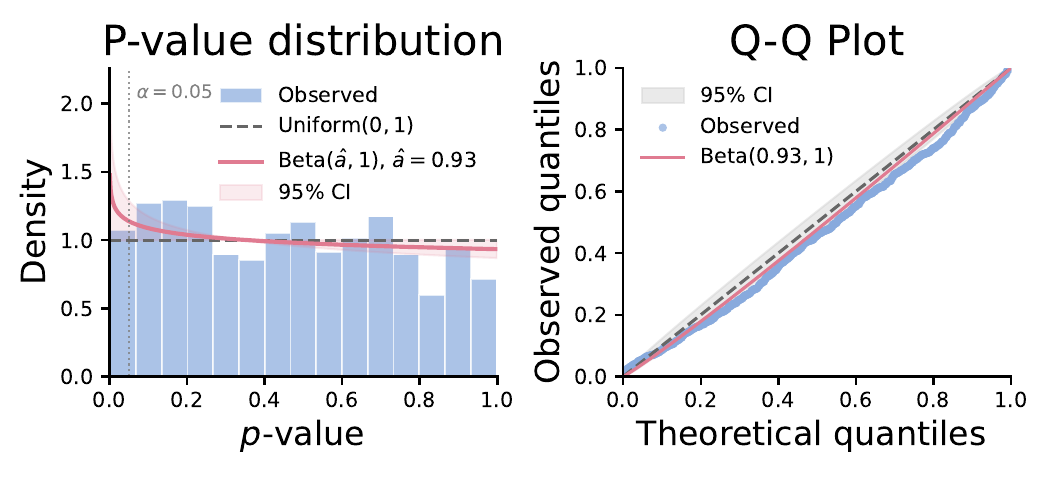}
\caption{ (Left) A histogram of observed p-values (blue) with the theoretical uniform density (dashed gray line) and the fitted $\mathrm{Beta}(\hat{a}, 1)$ distribution overlaid (red) using the gennorm kernel. The shaded region represents the 95\% credible interval derived from the posterior uncertainty in $\hat{a}$.
 (Right) A Q--Q plot comparing observed p-values to expected quantiles from a uniform distribution (dashed gray line) and from the fitted $\mathrm{Beta}(\hat{a}, 1)$ model (red). The gray shaded region denotes the 95\% confidence band for order statistics under the null. Systematic deviations from the uniform reference line indicate a departure from the null distribution, consistent with the fitted model.
}\label{fig:synth_pdist}
\end{figure}

\subsubsection*{Synthetic Data}

\begin{table}[ht]
\centering
\begin{tabular}{lcc}
\textbf{Distribution} 
 & \textbf{Alpha estimate} & \textbf{Clopper-Pearson}
 \\
\midrule
Gaussian & $0.49\;(0.46,0.53)$ & $0.23$ \\
Hypersecant & $21.1\;(19.8,22.0)$ & $0.0056$ \\
Student-t & $0.71\;(0.67,0.76)$ & $0.11$  \\
Laplace & $0.37\;(0.34,0.39)$ & $0.39$ \\
Gennorm & $0.64\;(0.60,0.68)$ & $0.13$  \\
\midrule
\end{tabular}
\caption{Effect of mixture distribution on conservativeness}
\label{tab:synthetic_lmr}
\end{table}

\subsubsection*{Electrophysiology Data}

\begin{table}[ht]
\centering
\begin{tabular}{l|cc|cc}
 & \multicolumn{2}{c|}{\textbf{Conditions Averaged}} 
 & \multicolumn{2}{c}{\textbf{Conditions Separated}} 
 \\
\textbf{Experiment} 
 & \textbf{Significance Prob} & \textbf{Avg P-Value}
 & \textbf{Significance Prob} & \textbf{Avg P-Value}
 \\
\midrule
Social Preference & 1.0 & 0.043 & 1.0 & 0.043 \\
Tail Suspension & 0.75 & 0.047 & 0.0 & 0.56 \\
Both & 1.0 & 0.013 & 1.0 & 0.005  \\
\midrule
\end{tabular}
\caption{Batch effect detection in the electrophysiology data using tail suspension data as training set.}
\label{tab:electrophysiology_comparison_tst}
\end{table}

\begin{table}[ht]
\centering
\begin{tabular}{l|cc|cc}
 & \multicolumn{2}{c|}{\textbf{Conditions Averaged}} 
 & \multicolumn{2}{c}{\textbf{Conditions Separated}} 
 \\
\textbf{Experiment} 
 & \textbf{Significance Prob} & \textbf{Avg P-Value}
 & \textbf{Significance Prob} & \textbf{Avg P-Value}
 \\
\midrule
Social Preference & 1.0 & 0.014 & 0.63 & 0.048 \\
Tail Suspension & 0.0 & 0.67 & 0.0 & 0.12 \\
Both & 1.0 & 0.013 & 1.0 & 0.005  \\
\midrule
\end{tabular}
\caption{Batch effect detection in the electrophysiology data using social data as training set.}
\label{tab:electrophysiology_comparison_soc}
\end{table}

\section{Comparison with the LMR Method}
\label{app:additional_results}

We now compare the results of our bootstrap method with the method advocated in~\citet{lo2001testing}. 

\begin{table}[ht]
\centering
\begin{tabular}{l|cc|cc}
 & \multicolumn{2}{c|}{\textbf{Lo-Mendell-Rubin Adjustment}} 
 & \multicolumn{2}{c}{\textbf{Bootstrap Calibration}} 
 \\
\textbf{Distribution} 
 & \textbf{Alpha estimate} & \textbf{Clopper-Pearson}
 & \textbf{Alpha Estimate} & \textbf{Clopper-Pearson}
 \\
\midrule
Gaussian & $0.49\;(0.46,0.53)$ & $0.22$ & 0.63 & 0.048 \\
Hypersecant & $20.0\;(18.7,21.2)$ & $0.0059$ & 0.0 & 0.12 \\
Student-t & $0.69\;(0.65,0.74)$ & $0.11$ & 1.0 & 0.005  \\
Laplace & $0.37\;(0.34,0.39)$ & $0.39$ & 1.0 & 0.005  \\
Gennorm & $0.63\;(0.59,0.67)$ & $0.14$ & 1.0 & 0.005  \\
\midrule
\end{tabular}
\caption{A comparison of LMR vs Bootstrap on synthetic data corresponding to liquid biopsy data.}
\label{tab:lmr_vs_bootstrap}
\end{table}

\end{document}